 \documentclass[manuscript]{aastex}

\newcommand{\etal}{\rm {\it et al.}}

\usepackage{lineno}
\linenumbers*[1]

\shorttitle{Temporal Deconvolution of  GRB Light curves}
\shortauthors{P.~N. Bhat \etal}

\begin{document}


\title{Temporal Deconvolution study of Long and Short Gamma-Ray Burst Light curves}

\author{P.N. Bhat\altaffilmark{1}, Michael S. Briggs\altaffilmark{1},  Valerie Connaughton\altaffilmark{1}, Chryssa Kouveliotou\altaffilmark{2}, Alexander J. van der Horst\altaffilmark{4}, William Paciesas\altaffilmark{1}, Charles A. Meegan\altaffilmark{4}, Elisabetta Bissaldi\altaffilmark{3}, Michael Burgess\altaffilmark{1}, Vandiver Chaplin\altaffilmark{1}, Roland Diehl\altaffilmark{6}, Gerald Fishman\altaffilmark{2}, Gerard Fitzpatrick\altaffilmark{9}, Suzanne Foley\altaffilmark{6}, Melissa Gibby\altaffilmark{7}, Misty M. Giles\altaffilmark{7}, Adam Goldstein\altaffilmark{1}, Jochen Greiner\altaffilmark{6}, David Gruber\altaffilmark{6}, Sylvain Guiriec\altaffilmark{1}, Andreas von Kienlin\altaffilmark{6}, Marc Kippen\altaffilmark{8}, Sheila McBreen\altaffilmark{9}, Robert Preece\altaffilmark{1}, Arne Rau\altaffilmark{6}, Dave Tierney\altaffilmark{9} and Colleen Wilson-Hodge\altaffilmark{2}}
\altaffiltext{1}{University of Alabama in Huntsville, NSSTC, 320 Sparkman Drive, Huntsville, AL 35805, USA}
\altaffiltext{2}{Space Science Office, VP62, NASA/Marshall Space Flight Center, Huntsville, AL 35812, USA}
\altaffiltext{3}{Institute of Astro and Particle Physics, University Innsbruck, Technikerstr. 25, 6020 Innsbruck, Austria}
\altaffiltext{4}{Universities Space Research Association, NSSTC, 320 Sparkman
 Drive, Huntsville, AL 35805, USA}
\altaffiltext{6}{Max-Planck-Institut f\"ur extraterrestrische Physik (Giessenbachstrasse 1, 85748 Garching, Germany)}
\altaffiltext{7}{Jacobs Technology, Inc., Huntsville AL 35806, USA}
\altaffiltext{8}{Los Alamos National Laboratory, PO Box 1663, Los Alamos, NM
 87545, USA}
\altaffiltext{9}{School of Physics, University College Dublin, Belfield, Stillorgan Road, Dublin 4, Ireland}


\begin{abstract}
The light curves of Gamma-Ray Bursts (GRBs) are believed to result from 
internal shocks reflecting the activity of the GRB central engine. Their
temporal deconvolution can reveal
potential differences in the properties of the central engines in the
two populations of GRBs  which are believed to originate from the deaths of 
massive stars
(long) and from mergers of compact objects (short).
We present here the results of the temporal analysis of 42 GRBs
detected with the Gamma-ray Burst Monitor onboard the Fermi Gamma-ray Space 
Telescope. We deconvolved the profiles into pulses, which we fit with lognormal functions. The distributions of the pulse shape parameters and intervals between neighboring pulses are distinct for both burst types and also fit with lognormal functions. We have studied the evolution of these parameters in different energy bands and found that they differ between long
and short bursts.  We discuss the implications of the differences in the temporal 
properties of long and short bursts within the framework of the internal shock model for GRB prompt emission. 
\end{abstract}





\section{Introduction}

The temporal structure of GRB light curves exhibits very diverse morphologies,
from single pulses to extremely complex multi-pulse structures. As a result,
morphological GRB classification attempts have not been successful and the only
established division of bursts into classes with different temporal
characteristics is based on their durations \citep{kmf93}.  The latter have
been found to distribute bimodally, with over 75\% of the events belonging in
the long class ($>2$\,s) when durations are measured in the 50-300 keV range. 
Since 1993 the GRB durations are mostly measured by
their $T_{90}$ ($T_{50}$) intervals, the times during which 90\% (50\%) of
the total event counts (or fluence) are collected \citep{kmf93}.
\citet{mcb94} and later \citet{hor02}, showed that durations ($T_{90}$) of both
long and short GRBs follow lognormal distributions separately. Several authors
have studied the deconvolution of GRB light curves into their constituents,
and have shown that in general, these are discrete, often overlapping
pulses with durations ranging from a few milliseconds to several seconds and
almost always asymmetric shapes, with faster rises than decays \citep{nor96,hak11}.
These highly varied GRB temporal profiles are suggestive of a stochastic
process origin.

Two distinct mechanisms have been proposed to explain the origin of pulses in
GRBs. In the external shock model, radiation pulses are emitted when a
relativistic shell ejected by the GRB central engine is decelerated by the
circum-burst material \citep{mes93}. A homogeneous medium leads to a single
pulse but an irregular, clumpy environment can produce a complex profile if a
large number of small clouds are present \citep{der99}. According to the
internal shock model \citep{rmm94}, the central engine generates relativistic
shells with highly non-uniform distribution of Lorentz factors and the pulses
are formed by the collision between a rapidly moving shell with a slower shell.
Thus in principle the variability of the GRB light curves may directly
correspond to the activity of their central engines \citep{fdm03, enp02}.
Hence the studies of pulse properties are important to determine whether
GRB sources require engines that are long lasting or impulsive \citep{der04}.

Investigations linking GRB properties with their pulse characteristics have
already been carried out by several authors
\citep{nor96,lif96,qui99,lee00,mcb01,hak08}. \citet{nor96} were the first to
deconvolve the profiles of long and bright GRBs detected with the Burst And
Transient Source Experiment (BATSE) onboard the Compton Gamma Ray Observatory
({\it CGRO}) into pulses and study the pulse shape parameters as a function of
energy. \cite{vgp00} were the first to fit lognormal functions to pulses in
short GRBs detected with BATSE. \cite{mcb01} applied a pulse identification
algorithm on a set of BATSE short bright bursts and derived their pulse shape
parameters; they concluded that the pulse rise and decay times follow lognormal
distributions. However, the BATSE GRB light curves used in these studies had a
time resolution of 64 ms for long bursts, which could have masked narrower 
pulses in those bursts. However the short burst studies have been carried out 
using higher resolution data.

A long standing question has been, therefore, whether the representative time
scales associated with pulses of long GRBs form a separate class from those in
short bursts perhaps reflecting the two different prevalent models for their
origin, i.e., long bursts originate from
the collapse of massive stars \citep{woo06a,woo06b}, while short GRBs result
from the merger of two compact objects \citep{eic89,nar92}. If we could
deconvolve the GRB light curves in terms of simpler pulse shapes, we could
potentially identify the differences in the central engines of long and short
GRBs. In this paper, we show that there is a fairly high degree of determinism
underlying the complex nature of the GRB temporal profiles. In \S~2 we
describe the instrument and the selection criteria for our sample, and in
\S~3 we expand on our analysis technique. In \S~4 we decompose the
high-time resolution GRB data of the Gamma-ray Burst Monitor (GBM) onboard the
{\it Fermi} Gamma-ray Space Telescope (hereafter {\it Fermi}) into individual
pulses  and examine the distributions of the pulse shape parameters for long
and short duration GRBs. Further we apply the same analysis technique to GRB
light curves in various energy bands and study the pulse shape evolution with
energy. We discuss our results in \S~5. 

\section{Instrumentation and Data Selection}

GBM is an uncollimated all-sky (field of view $\geq$ 8 sr) monitoring
instrument. It consists of an array of 12 NaI(Tl) scintillation detectors
mounted in clusters of three around the spacecraft. Each NaI(Tl)
detector is 12.7\,cm in diameter by 1.27\,cm thick, and covers an energy range
from 8 keV to 1 MeV. In addition, GBM includes two Bismuth Germanate (BGO)
detectors, each 12.7 cm in diameter by 12.7 cm thick, placed on either side of
{\it Fermi}. The BGOs cover energies above 150 keV up to a maximum of 40 MeV 
\citep{mlb09}.

The GBM on-board software incorporates burst triggering on time scales as 
short as 16 ms. All triggers generate time-tagged
event data (TTE) consisting of the photon arrival time and energy as 
deposited from each of the 14 detectors with a temporal resolution
of 2~$\mu$s \citep{mlb09}. The very high temporal resolution and large energy 
band-width are major assets for the study of GRBs in general and the
study of short events, in particular. A pre-burst ring buffer saves about 
half a million events before the trigger, which corresponds to a time interval
of~$\sim$30 seconds. The TTE data are produced for $\sim$300 seconds after the
trigger. All short bursts and a bulk of the long bursts have full temporal 
coverage by TTE data. The energy range for both NaI and BGO detectors is
digitized into 128 channels, pseudo-logarithmically spaced to provide channel
widths less than each detector energy resolution up to 12 MeV though TTE data 
are available at coarser resolution up to 40 MeV. During the 3 years since its
launch (2008 June 11) GBM has collected over 700 GRBs. During the first year 
GBM detected 225 GRBs of which 59 were BGO bright bursts \citep{bis11}.
This is to ensure the burst is sufficiently hard to allow pulse decomposition
analysis in different energy channels. Out of the latter dataset we chose long
bursts with the product of fluence and peak flux 
(1.024s) values greater than $1.0\times 10^{-4}$ and $5.0\times 10^{-6}\,{\rm erg~ph/cm^4/s}$
for long and short bursts respectively. Burst fluences and peak fluxes estimated
in the energy range 10-1000\, keV are taken from the GBM Gamma-ray burst catalog
\citep{pac11}. As a result, the final sample includes 32 long bursts
with fluences ranging from $5.5\times 10^{-6}$ to $2.7\times 10^{-4}\,
{\rm erg/cm^2}$ and 10 short bursts with fluences 
ranging from $8.5\times 10^{-7}$ to $8\times 10^{-6}\,{\rm erg/cm^2}$.
This unusual selection criteria is simply to eliminate weak and long bursts with
fluences above the threshold that are difficult for the pulse decomposition analysis.
We have also used the BGO data in the current analysis. The burst durations 
and the number of fitted pulses are listed in Table \ref{GRB_table}.
Possible selection effects arising from our choice of burst sample is
assumed to be small in the present analysis.

For each burst we summed the TTE data of the four NaI detectors that registered
the highest gamma-ray signal
(with an angle  to the burst direction of $\leq$ 60$\degr$) to derive their 
light curves with a resolution of 1\,ms. In the case of BGO detectors, the light
curves from both the detectors were summed. Each light curve included
the entire burst and background regions up to about 10-20 s before and after the
burst. We varied the temporal resolution used for the analysis depending on the
burst intensity (see next section for details). The analysis described below was
performed on the entire energy range of NaI ($8-1000$ keV) and BGO ($0.15-45.0$ MeV)
detectors  as well as in six NaI energy ranges per burst ($8-520$ keV;
see also Table \ref{Tab1}). It may be noted that there are uncertainties in the
energy edges listed in Table \ref{Tab1} that arise primarily from the finite energy
resolution of the GBM detectors \citep{mlb09}.

\section{Analysis Technique}

In general, a parameter which can be written as a product of $\geq$ 3 random
variables tends to follow a lognormal function \citep{ajb69}. Since the
pulse shape parameters of GRB light curves can be described as such a product, 
we were motivated to use a similar procedure to test this hypothesis \citep{iok02}.
\citet{iok02} argue that the distribution of a product of variables tends to
the lognormal distribution as the number of multiplied variables increases,
the distribution of pulse width may be closer to the lognormal distribution
than that of the pulse intervals between successive pulses.
It has already been shown that the long and short GRB durations, the time 
interval between successive pulses \citep{mcb94}, fluence and pulse intervals between successive pulses
within each burst \citep{lif96}, pulse durations \citep{enp02a} and spectral
break energies \citep{pre00} do follow lognormal distributions. 

We also choose a lognormal function to fit pulses in a GRB light curve.
A lognormal function in this case has 4 free parameters, namely the 
amplitude (A), mean ($\mu$), standard deviation ($\sigma$) and time. The 
advantage of choosing this functional form is that it
converges in all cases even when the shape of the light curve is very complex,
where the pulses are often overlapping. The pulse shape parameters are derived
from the fit parameters using the following formulations. 
 
A lognormal function is represented as:
 
$$f(x) =\left\{\begin{array}{ll} \frac{A}{\sqrt{2\pi}x\sigma}
\exp\left[-\frac{(\log~x-\mu)^2}{2\sigma^2}\right] & {\rm
if}~ x>0\\
&\\ 0 & {\rm if}~ x\leq 0\\\end{array} \right.$$
\noindent
where, $\mu$ and $\sigma$ are the sample mean and standard deviation of 
$\log{x}$, and A is the
amplitude. The rise time, $\tau_r$, decay time, $\tau_d$, and the full width
at half maximum of each pulse, FWHM, can be derived from the fit parameters of 
the lognormal function given above. The rise and decay times are measured from
the time differences at 10\% and 90\% of the peak amplitude of a pulse.

$$\tau_r = \exp{(\mu-\sigma^2)}\left[ \exp \left( -\sigma\sqrt{2\log\left( \frac { 10}{9}\right)}\right) - \exp\left( -\sigma\sqrt{2\log{(10)}} \right)\right] $$

$$\tau_d = \exp{(\mu~-~\sigma^2)}\left[ \exp\left( \sigma\sqrt{2\log {(10)}}\right) - \exp\left( \sigma\sqrt{2\log\left( \frac{10}{9} \right)} \right) \right] $$

$$ FWHM = \exp{(\mu~-~\sigma^2)}\left[\exp\left( \sigma\sqrt{2~\log{(2)}} \right) - \exp\left( -\sigma\sqrt{2~\log{(2)}} \right) \right] $$

\begin{figure}
\epsscale{.80}
\plotone{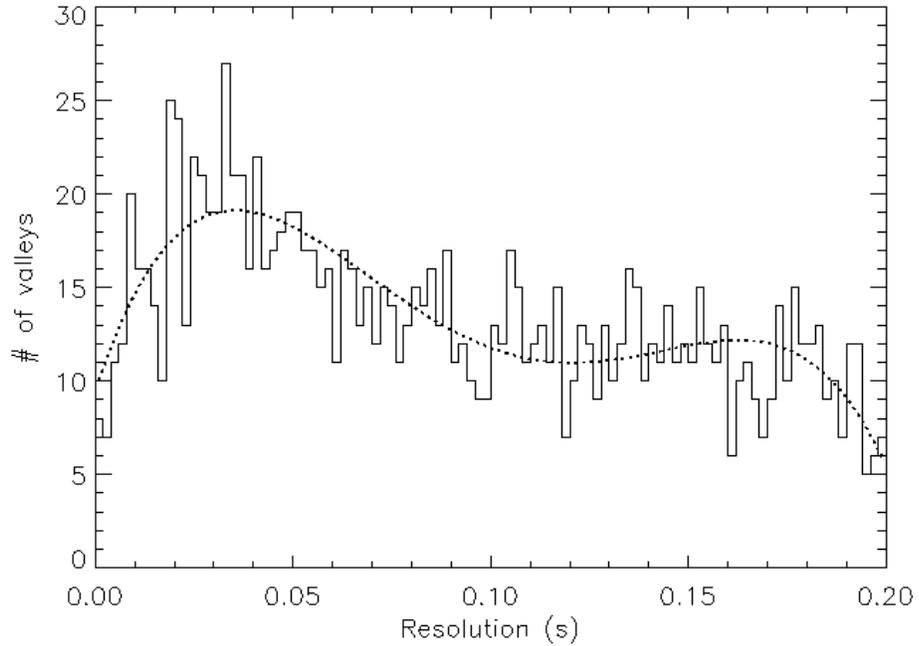}
\caption{A sample histogram of the number of valleys identified by an algorithm based on
the method suggested by \citet{lif96}, as a function of bin-width of the light
curve for GRB\,080723D. The number of valleys increases initially and reaches a
broad maximum at the optimum bin-width (25-50\,ms)and then gradually falls at very coarse resolution. 
The curve is a ploynomial fit to guide the eye only. The number of valleys and the
bin-width at the maximum agrees with that chosen for this GRB which is 18 valleys at a
time resolution of 50 ms.
\label{valley_res}}
\end{figure}

For each GRB we initially selected the number of possible pulses contained in
the light curve by visually identifying the significant valleys on either side
of a pulse. This process was repeated for each burst varying the temporal
resolution of the summed light curve until the number of valleys reached a
maximum. If the resolution was too fine, the pulses were burried in statistical
fluctuations and hence the number of identified valleys was too small. At very
coarse resolutions the closely spaced pulses merged with each other also
resulting in a reduced number of valleys. The number of valleys is maximum at
the optimum temporal resolution for a given burst. Figure \ref{valley_res}
shows such a histogram where the number of valleys identified automatically by a routine based
on the technique of \citet{lif96}, as a function of bin-width of GRB light curve.
The number of valleys increases initially with increasing binwidth and then
reaches a broad maximum at a resolution in the range 25-50\,ms and then
falls gradually with further increase in the bin-width. The number valleys estimated
manually for this burst was 18 at a chosen optimum bin-width of 50 ms which agrees well
with those estimated objectively. The mean time resolution for all the GRBs in our sample is $\sim$\,40 msec. 

The array of valleys was then used as input to the pulse fitting routine.
It generates initial guesses of the amplitudes, means and the standard
deviations based on the number of counts in the light curve between a pair of valleys 
while the pair of valleys are used to estimate the initial guess of time parameter. 
The routine then simultaneously fitts lognormal functions to pulses at optimum times
and a quadratic to the background. It compared the model light curve with the data
and minimized its $\chi^2$ value by varying the pulse shape parameters and the
position of the pulses. The goodness of fit, $n$, was finally calculated by
computing the likelihood parameter as {-2$\ln \pounds$} (which approaches
Pearson's $\chi^2$ for large model values) divided by the number of degrees of
freedom (dof). The number of dof is the difference between the number of data
points in the light curve minus the total number of fitted parameters. This
procedure was then repeated for the light curves in the first four to six energy
bands (depending on the burst intensity in the higher energy bands) 
shown in Table \ref{Tab1}, defined so that a typical GRB light curve had similar
signal to noise ratio in each channel. After pulse fitting we
used the pulse mean positions to compute the intervals between successive pulses, while 
the variation of the pulse shape parameters in different energy bands were 
used to study the spectral evolution of the pulse shapes. The pulse mean positions 
refer to the mean times(with respect to the trigger time) of the lognormal pulses.

To test the integrity of the fit we computed a weight for each of the fitted
pulses in a light curve by estimating the percentage change in the goodness of
fit parameter with that pulse excluded. Pulses with weights less than 2\% were
excluded from the fit as they most likely were due to statistical fluctuations.
The overall goodness of fit did not change more than 10\% compared to its value
before any pulse rejection. No case was found where an additional pulse was
needed to improve the residuals. Thus the pulse fitting procedure was optimized
to ensure removal of spurious pulses.

To check the robustness of our fits, we also performed a series of simulations 
as follows. We chose a set of pulses fitted to a light curve and generated a
synthetic light curve using these pulses superposed over the burst fitted
background. We then reduced the light curve intensity in steps of 10\%, starting
at 100\%, and added statistical noise to each bin. Each light curve was fitted
by the normal procedure and recovered entirely until the intensity was decreased
to 50\%. The degree of percentage recovery declined thereafter, and reached 75\%
of the original, when the intensity was reduced to 10\% of the total. We
concluded that the fit is robust for a large range of burst intensities.

Further we reduced the duration of the simulated burst by a factor and fitted 
the light curve with the lognormal functions as before. Each time the 
separation 
between the pulses too reduced by the same factor. Hence there was no lower
limit on the inter-pulse separation caused by the closeness of the successive 
peaks in the light curve. This was tested by reducing the burst duration by a
factor of 1000.
 
Finally, to address the issue of the interdependency of the rise and decay times of
a lognormal function we tried two new functions, where these times can vary independently.
They are:

$f_1(x) =\frac {A}{(\tau _d - \tau _r)}\left[\exp(-\frac{x}{\tau _r})-\exp(-\frac{x}{\tau _d})\right]$, 
\noindent
\citep{leo94}

$f_2(x) = A\lambda\, \exp(-\frac{\tau _1}{x} - \frac{x}{\tau _2})\  {\rm for\ } x\, >\, 0 $,
\noindent
\citep{nor05}

\noindent
where $\lambda = \exp(2\mu)$
\noindent
and
$\mu = \left(\frac{\tau _1}{\tau _2}\right)^{\onehalf} $.

We find that for long bursts with a small number of pulses, both the 
above functions fit the
light curve as well as the lognormal function. However, in case of bursts with complex
light curves consisting of several overlapping narrow pulses, these
functions fit poorly resulting in large $\chi^2$ compared to a fit with a
lognormal function. Since the present analysis aims at a comparison of the 
pulse properties of long and short bursts, we opted to use the lognormal 
function, which describes the light curve best in all cases. 

\section{Results}

\begin{figure}
\epsscale{.80}
\plotone{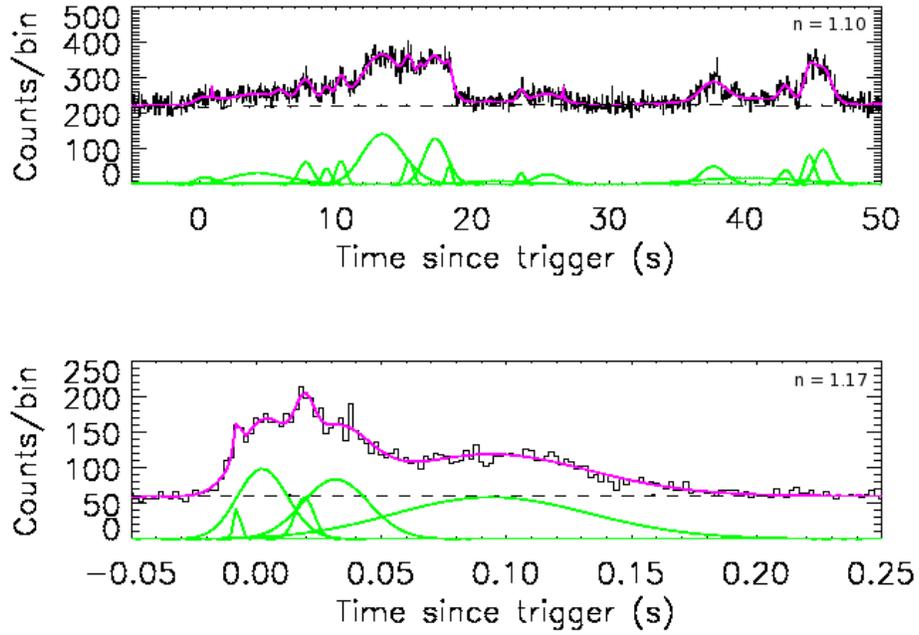}
\caption{A sample pulse fit to one long burst GRB\,080723D (upper plot) and 
one short burst GRB\,090227B (lower plot). The histogram in black is the GRB 
light curve and the fitted background is shown as black dashed line. The pulses 
shown in green are the lognormal pulses fitted to those in the light curves. 
The sum of the background model and the fitted pulses is shown as purple 
continuous line.  The goodness of fit parameter, $n$, is indicated at the top 
right corner of each plot.
\label{sample_fits}}
\end{figure}

\begin{figure}
\epsscale{.80}
\plotone{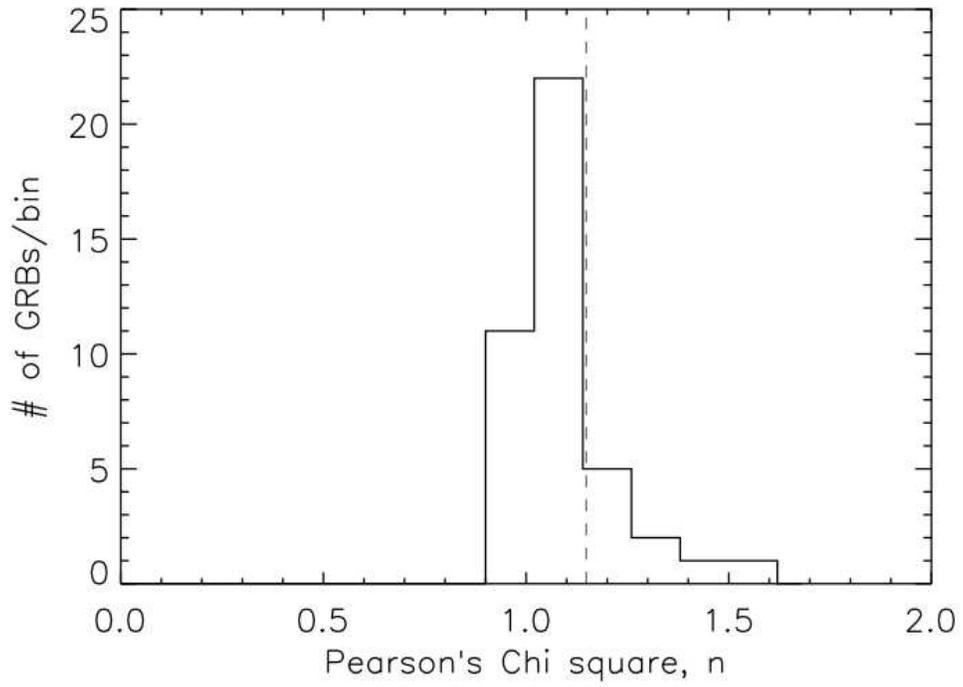}
\caption{A distribution of the goodness of fit parameter n, viz. Pearson's 
chisquare estimated from the liklihood ratio for each fit. The vertical dashed 
line shows the mean value of the entire sample.
\label{chi_dist}}
\end{figure}

We performed the analysis described above on all 42 GRBs in our sample. 
The pulse shape parameters from the analyses of the entire sample of light curves used here
are summarized in 2 tables which are available in the online version of this paper.
Figure \ref{sample_fits} shows an example of fitted pulses to light curves of one long
(GRB\,080723D, upper plot) and one short (GRB\,090227B, lower plot) GRB. The 
quality of each fit, $n$, is
indicated at the top right hand corner of each panel. The mean value of this
parameter for all the 42 fits is 1.15 with a standard deviation of 0.13. Figure
\ref{chi_dist} shows the distribution of $n$. We note that it peaks around 1, as
expected since $n$ is expected to follow the $\chi^2$ distribution.

\subsection{Pulse shape parameters}

\begin{figure}
\epsscale{.80}
\plotone{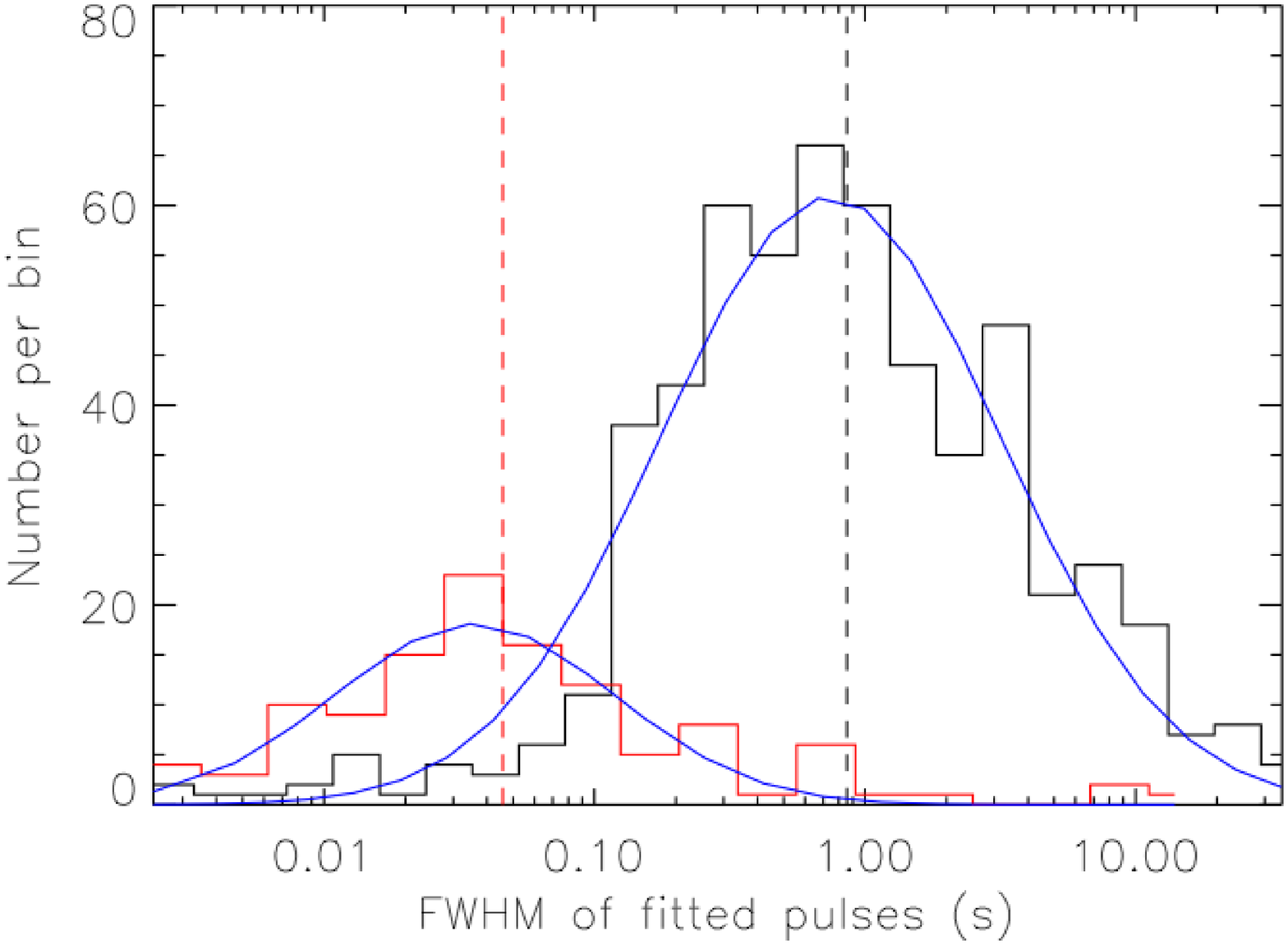}
\caption{Distributions of the pulse widths (FWHM) for long (histogram shown in black) and short 
bursts (histogram shown in red). A lognormal function is fitted to each of the distributions.
The mean values of FWHM (from the fit) for long and short bursts  are 0.89\,s and 0.055\,s and the 
standard deviations are 5.2\,s and 4.6\,s respectively.
The vertical dashed lines are the median values of FWHM for each class of GRBs.
\label{GRB_fwhm}}
\end{figure}

For each burst, we derived $\tau_r$, $\tau_d$ and $FWHM$ for every fitted pulse
from the formulae listed in the previous section. 
Figure
\ref{GRB_fwhm} shows the distributions of pulse FWHM for long and short bursts
also independently fitted with lognormal functions with mean values of pulse widths of 0.95\, s and 
0.06\, s, respectively.  The distributions are overlapping but distinct for the 
two types of bursts.  We note that short burst light curves consist of
distinctly narrower pulses compared to long GRBs.

\begin{figure}
\epsscale{.80}
\plotone{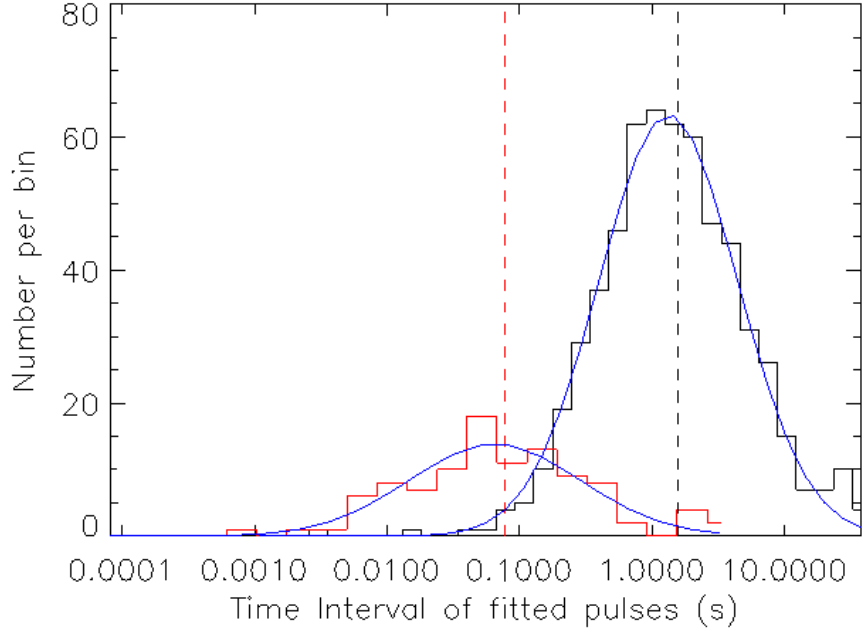}
\caption{Distributions of the time intervals between successive pulses ($\Delta t$) for 
long (histogram shown in black) and 
short (histogram shown in red) bursts. A lognormal function is fitted to each of the distributions. 
The mean values of $\Delta t$ (from the fit) for long and short bursts are 
1.53 s and 0.076\,s and the standard deviations are 3.6 and 5.1 respectively.
The vertical dashed lines indicate the median values of the time intervals between successive
pulses for each class of GRBs.
\label{GRB_int}}
\end{figure}

The pulse widths and intervals between successive pulses are primary attributes which could
ultimately reveal important clues about GRB physics. Figure \ref{GRB_int} shows
the distributions of the time intervals ($\Delta t$) between adjacent
pulse positions for long and short GRBs. Again we fit the distributions independently with lognormal functions.
The means of the fitted lognormal functions are 1.6\,s and 0.08\,s for long and short bursts,
respectively. The pulses in short bursts are about 20 times more closely spaced
than those in the long bursts. The range of intervals between successive pulses spans nearly 3 decades
both for long and short GRBs consistent with the earlier results for BATSE
bursts \citep{nor96}. An exponential function does not fit the cumulative
distributions of the intervals between successive pulses well, indicating that most likely the GRB 
pulses do not follow a Poisson distribution in time.

It may be noted that the average redshift of short bursts is smaller than that
of long bursts (see section 5).  Hence the average pulse widths and and intervals between successive pulses of
long bursts could be larger by a factor of 1.74 because of this effect. However
the redshift effect is too small to account for the separation between them as seen in 
figures \ref{GRB_fwhm} and \ref{GRB_int}.

\begin{figure}\epsscale{.80}
\plotone{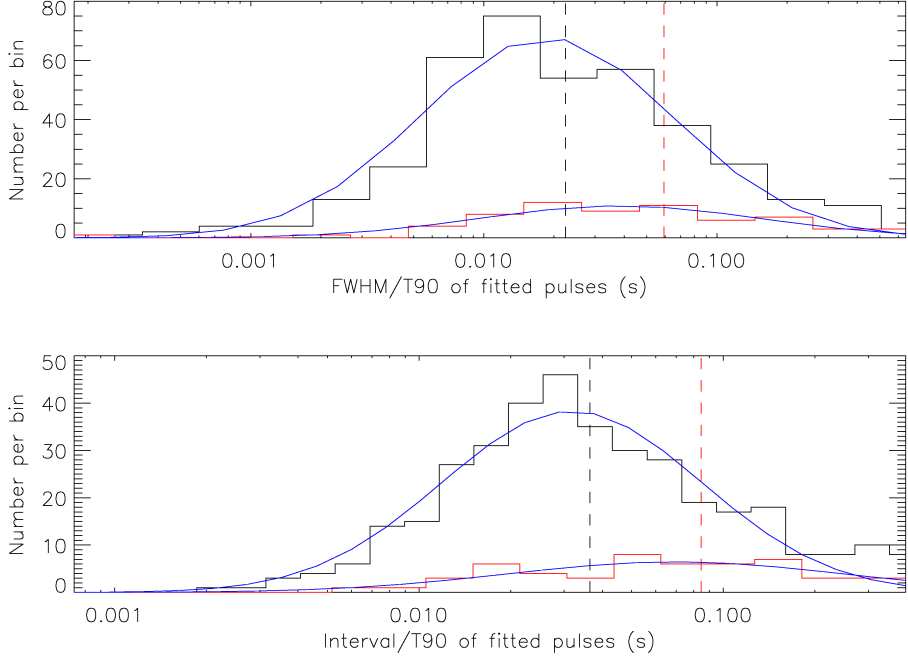}
\caption{Distributions of the ratio of pulse widths (FWHM) of a burst to
its total duration, $T_{90}$, for long bursts (histogram in black) and short
bursts (histogram in red, upper panel). Each of the histograms is fitted with
a lognormal function (shown as continuous curves in blue). The vertical dashed lines show
the median values for long and short bursts respectively. The lower panel shows
similar distributions of the ratio of the GRB time interval between successive pulses and the burst
duration ($T_{90}$). Considering large 
errors in the short burst durations the two distributions may be considered to
be consistent with each other in both the cases. The lag between the two
distributions are $1.0\pm 0.7$ (upper plot) and $2.0\pm 0.7$ bins (lower plot)
\label{fwhm_dur}}
\end{figure}

To compare the pulse width contribution to the total duration in short and long
bursts, we derive the ratio of the pulse width (FWHM) to the $T_{90}$ of
each burst. Figure \ref{fwhm_dur} (top panel) shows the histograms of these 
ratios for long and short bursts. Also shown are the median values of these
distributions. The same figure (lower panel) shows similar distributions for
the ratio of the pulse time intervals between successive pulses and the burst durations again
for both burst types. The distributions for long and short bursts are 
overlapping and consistent with each other, considering large uncertainties
in the short burst durations (Table \ref{GRB_table}). In order to quantify
the degree of overlap we estimated the time lag between the two distributions
as follows. We estimate the cross-correlation coefficient (CC) as a function of lag
between the two histograms. It is found that maximum of the CC values are 0.94 and
0.91 respectively for the distributions implying that the corresponding distributions
for long and short bursts are correlated. The lag is defined as its value where the CCF peaks.
In addition, we estimated lag 100 times for each pair of histograms while adding Poisson
noise to the data each time. The standard deviation of the lags from the 
simulated histograms is the error on the estimated lag. The lags so estimated are
$1.0\pm 0.7$\,bins and $2.0\pm 0.7$\,bins for the two distributions respectively. 
The values of lags are close to zero supporting the general observation that the short 
GRBs are similar to long GRBs compressed in time \citep{gui10}.

We now compare the GBM results with those of the BATSE bursts. \citet{enp02a}
use a modified peak finding algorithm first reported by \citet{lif96}, to a
sample of 68 long BATSE bursts and report that the pulse durations follow a
lognormal distribution. The pulse interval distribution which peaks around an
interval of 1.0\,s also exhibits an
excess of longer intervals between successive pulses with respect to a lognormal function. This is
consistent with the analysis of 319 long bright BATSE GRBs by \cite{qui02}. 
They show that for long GRBs the distribution of intervals between successive pulses 
peak at 1\,s and intervals longer than 15\,s, which form 5\% of 
the total, do not fit the lognormal distribution. They also show that these intervals 
between successive pulses are consistent with a power law. The origin of this excess has been attibuted to
the existence of quiescent times between successive peaks. In the present 
data, the distribution of pulse intervals for long bursts peaks around 1\,s and 
the fraction of intervals between successive pulses above 15\,s is $(3.8\pm 0.8)$\% which is 
consistent with the above result.  The interval distributions for both long and
short bursts are best fit by lognormal functions (Figure \ref{GRB_int}). 
The lognormal fit for long bursts shows a hint of an excess of long intervals between successive pulses even though
statistically not compelling because of smaller number of bursts in our sample.
 
\begin{figure}
\epsscale{.80}
\plotone{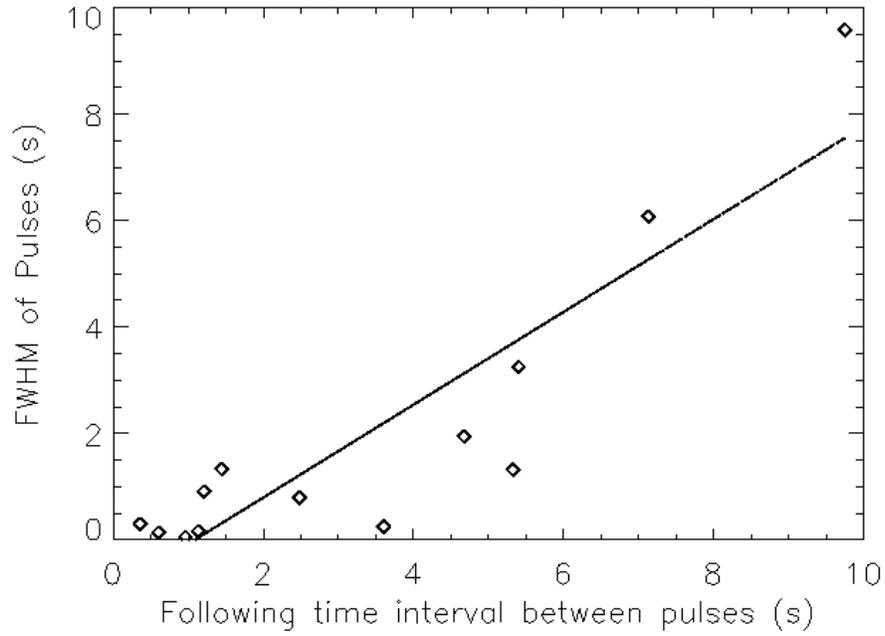}
\caption{A plot of the pulse FWHM as a function of following
time interval between successive separable pulses for a long burst GRB090626 which show 
a linear correlation coefficient of 0.896. The null hypothesis probability 
is $2.4\times 10^{-4}$ after taking into account of the number of GRBs searched (7 in this case).
The black dashed line shows a linear fit to the data points. 
\label{cc_plot}}
\end{figure}

\citet{enp02a} find a positive correlation ($>$ 70\%)between pulse width and the
preceding interval and a weaker correlation between pulse width and the
following time interval. They considered only bursts with more than 12 well separated
pulses and the total number of long bursts in their sample meeting this criteria was 12. 
There are 7 long bursts meeting these criteria in our sample. A search for such a 
correlation in our GRB sample has been carried out. In addition, 
we also searched for possible correlations between the pulse amplitude and the preceding or
following time interval.
We found one case of significant correlation (for GRB090626A) between the pulse width 
(FWHM) and the following time interval between successive pulses. 
The Pearson's linear correlation coefficient is 0.896 with a statistical null 
hypothesis probability of
$2.42\times 10^{-4}$. The corresponding plot is shown in Figure
\ref{cc_plot}. We found good correlations in a few other cases. However these 
correlations were found to be contributed by one or two
deviant points and hence likely to be spurious. 
No significant correlations were found between the pulse amplitude and the time
intervals in any burst in our limited sample. 
It seems that there are certain types of long bright GRBs 
which show such correlation between the pulse width and following time
interval between successive separable pulses. The implications of 
these correlations is not clear at present. 

\subsection{Spectral Evolution of Pulse Shape Parameters}

\begin{figure}
\epsscale{.80}
\plotone{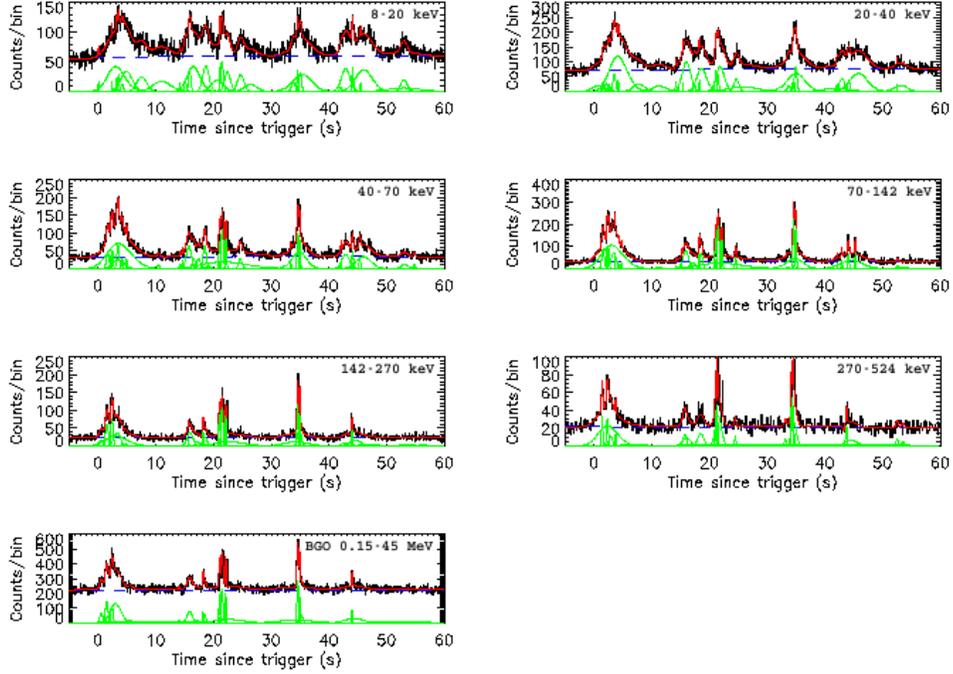}
\caption{Sample pulse fits to the lowest 6 energy channels of NaI and the
full energy range light curve from BGO detector of a long 
GRB 090626A. The histogram in black is the GRB light curve with a time
resolution of 50 ms. The pulses shown in green are fitted lognormals to those
in the GRB light curve. The horizontal dashed blue line is the fitted 
background while red curve is the fitted light curve resutling from summing 
the green and black curves. The temporal features in the light curve  and hence the
fitted pulses are narrower at higher energies. \label{pulse_fit_energy}}  
\end{figure}

Figure \ref{pulse_fit_energy} shows a set of light curves of GRB\,090626A 
in 7 different energy bands. Here we were restricted by the statistics in the
higher energy channels, which did not allow reliable pulse fitting beyond 524
keV for many bursts. The histograms of the pulse shape parameters were
generated as above in each energy range for short and long bursts. We assigned
a mean energy for each range estimated as the geometric mean of the energy
boundaries of each band. The energy range of each light curve is indicated on 
each plot. The bottom panel is from a fit to the BGO
light curve of the same burst in the entire BGO energy range. 
As in figure \ref{sample_fits} the individual pulses shown at the 
bottom of each panel when superposed on the quadratic background (shown as 
dashed line) describe the burst light cuve shown as continuous line in red.
Table \ref{Tab3} lists the number of pulses fitted for a smaple of long GRBs
in different energy bands. There does not seem to be a drastic change in the 
number of fitted pulses in the NaI energy range. The pulse fitting analysis 
of the BGO light curves in various energy bands were limited to very few bursts
and hence the results from the analysis of full energy light curves only are 
used here.

\begin{figure}
\epsscale{.80}
\plotone{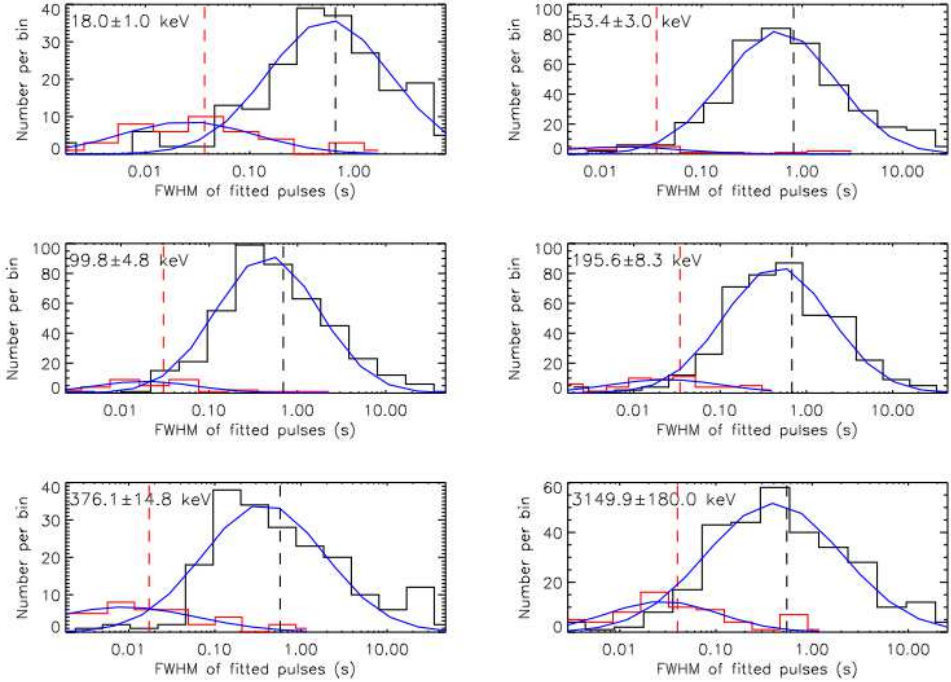}
\caption{Evolution of the distributions of pulse widths for long (histogram shown in black) 
and short (histogram shown in red)
GRBs as a function of energy. The first 5 distributions are from the NaI light 
curves in the first five energy bands as shown in Table \ref{Tab1}. Also shown
in the lowest right panel is a similar distribution for the total energy
light curve from the BGO detector. The histograms for long and short bursts are
fitted to lognormal functions and shown as continuous curves (in blue) in each energy band. 
The geometric mean
energies corresponding to each plot are indicated in each panel. The errors on
the mean energies are due to the finite energy resolution of GBM detectors.
\label{mean_fw_evoln}}
\end{figure}

Figure \ref{mean_fw_evoln} shows the distributions of pulse width (FWHM) of long
and short GRBs in different energy ranges. The distributions in each energy band
are well fit (shown as continuous curves) by lognormal functions. The width of
the fitted lognormals, as well as the values where the distributions peak, do 
not seem to change significantly with energy. The largest differences appear 
when we compare the two extreme energy bands, namely between 18\, keV and $\sim$
3.15\,MeV, with the latter widths being 0.04\,s and 0.5\,s narrower, for short 
and long GRBs, respectively.  

\begin{figure}
\epsscale{.80}
\plotone{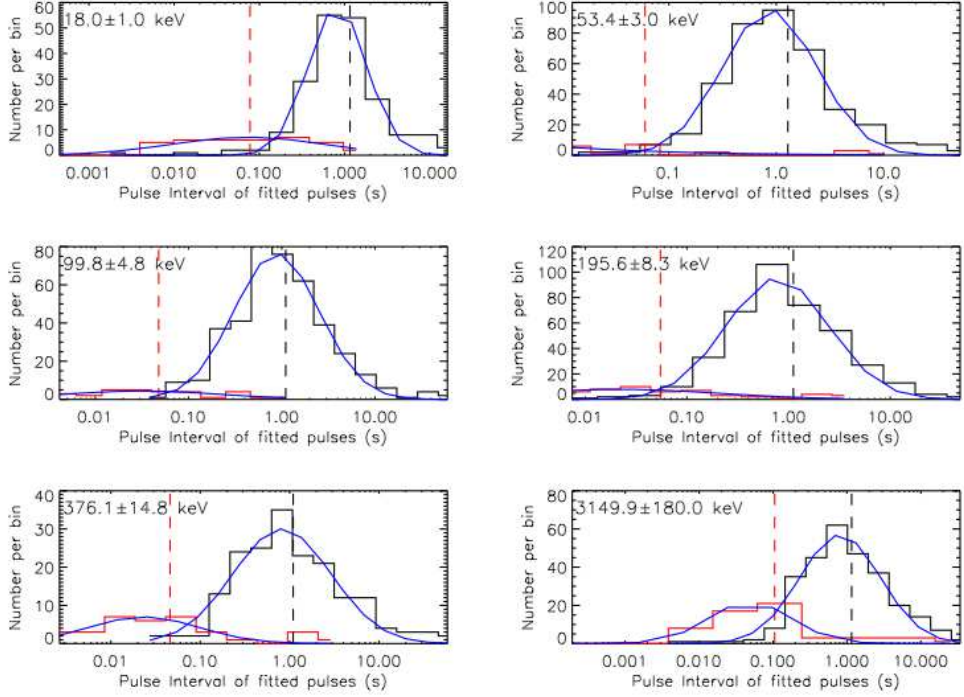}
\caption{Evolution of the distributions of pulse time intervals  between successive 
pulses for long (histogram shown in black) 
and short GRBs (histogram shown in red) as a function of energy. The first 5 distributions are from the 
light curves in the first five energy bands as shown in Table \ref{Tab1}.
Also shown in the lowest right panel is a similar distribution for the total
energy light curve from the BGO detector. The histograms for long and short bursts are
fitted to lognormal functions and shown as continuous curves (in blue) in each energy band. 
The geometric mean
energies corresponding to each plot are indicated in each panel. The errors on
the mean energies are due to the finite energy resolution of GBM detectors.
\label{mean_int_evoln}}
\end{figure}

\begin{figure}
\epsscale{.80}
\plotone{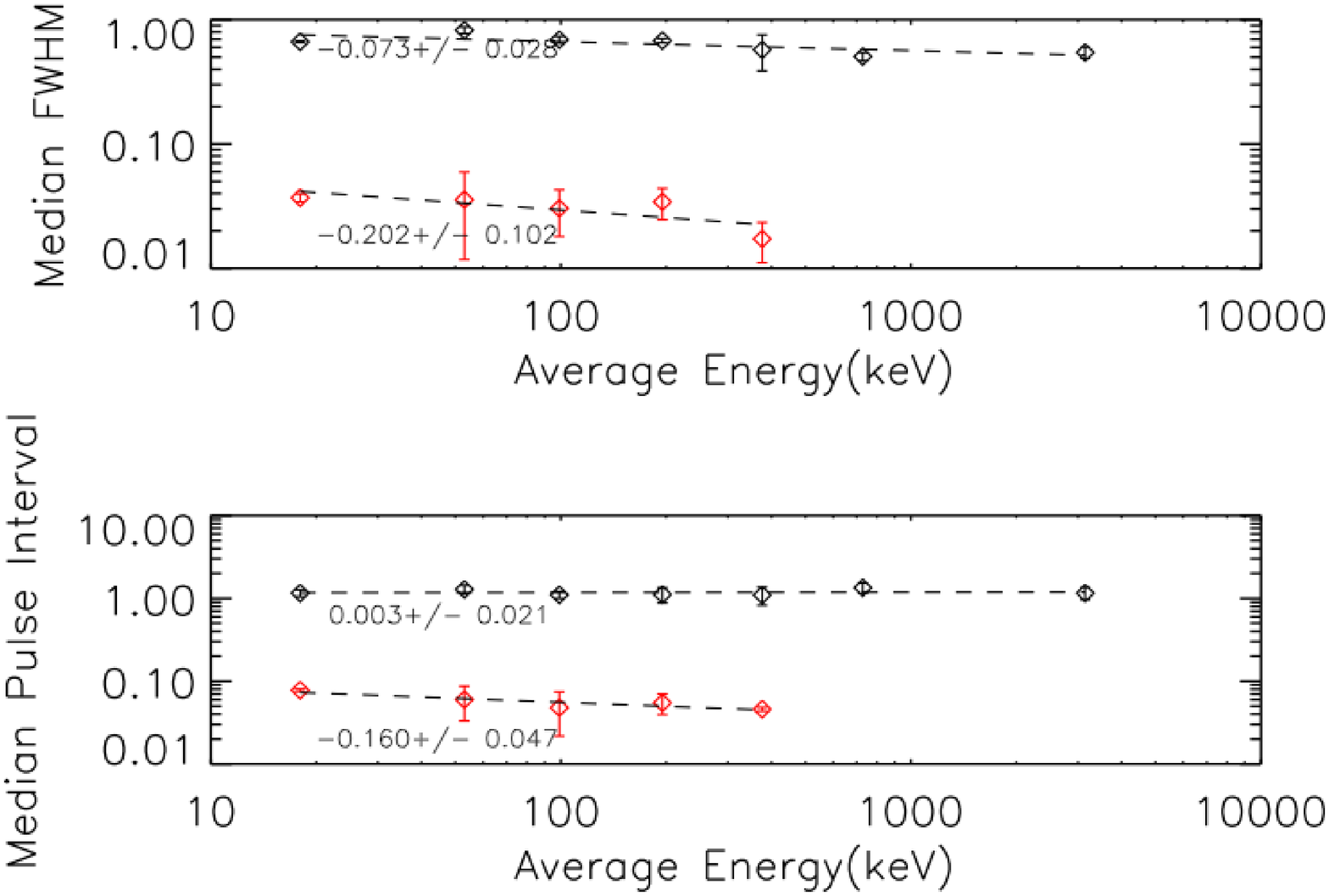}
\caption{Evolution of the median pulse width (top) and median pulse 
intervals between sucessive pulses (bottom) for long and short GRBs with energy.
The upper plot (in black) in each panel are the data for long bursts and the lower
plot (in red) are for short bursts.  In the case of short bursts
both the pulse width and pulse interval show a faster decrease with
increasing energy. In the case of long bursts on the other hand the
pulse width shows a slower decrease with increasing energy than that
for short bursts. The median pulse interval hardly seems to change with 
increasing energy in the case of long GRBs. The fitted power laws are shown
as dotted lines for each type of GRBs and the fitted slopes are indicated.
\label{mean_fw_int_evoln}}
\end{figure}

Figure \ref{mean_int_evoln} shows the evolution of the distributions of
time intervals between neighboring pulses of long 
and short bursts. Also shown are the best-fit lognormal functions in each 
energy band both for long and short GRBs. Figure \ref{mean_fw_int_evoln} shows
the variation of the median pulse width (FWHM) and median time interval between
successive pulses as a function of increasing energy for long and short bursts.
We note marked differences in the evolution of these 2 parameters for the two
types of bursts. In both cases the short bursts show a relatively rapid decrease
with energy as compared to long GRBs in agreement to earlier results where a
general tendency of GRB pulses to be narrower at higher energies has been
identified \citep{nor96}. The energy dependence of median pulse widths can be
represented as $\Delta t \propto ~E^{\alpha_{\rm w}} $ where
$\alpha _w = 0.07 \pm 0.03$ for long bursts, while $\alpha _{\rm w} = -0.2
\pm 0.1$ for short bursts. The median pulse interval also evolves very
differently in the case of long and short bursts. Both show a power law
dependence, with the exponents for long and short GRBs being
$\alpha _{\Delta \rm t} = 0.003 \pm 0.02$ and $\alpha _{\Delta \rm t} = -0.16
\pm 0.05$, respectively. The slope for long GRBs is consistent with zero,
indicating that the median interval size is constant with energy, while the 
short GRB pulses are more closely spaced at higher energies.



\section{Discussion}

Sari \& Piran (1997) argued that the observed temporal structure of a GRB
reflects the activity of the central engine that generates it. According to the 
internal shock model, the GRB pulses are formed by the collisions among 
relativistic shells ejected by the central engine with a distribution of 
Lorentz factors (${\gamma _e}$). 
A GRB pulse shape depends on three time scales. The hydrodynamic time scale, 
$t_{dyn}$ (that determines the pulse rise time), the angular spreading time 
scale, $t_{ang}$ (that determines the pulse decay time), and the cooling time 
scale, $t_{rad}$ (which is usually much shorter than the other two and can be 
ignored) \citep{kob97,kat97,fen96}. Hence the measured pulse shape parameters
have the potential to 
diagnose the pulse characteristics such as the bulk Lorentz factors, 
$\gamma_e$, shell radii and thicknesses \citep{koc03}.

Because of relativistic radiation-beaming only a small cone of opening angle 
${\gamma _e} ^{-1}$ is visible to the observer. The time difference between 
$\gamma-$rays emitted on-axis and off-axis constitutes the pulse decay. The 
off-axis $\gamma-$rays are delayed by $T _{ang} = \frac {R_e}{2{\gamma _e}^2}$,
where $R_e$ is the typical radius characterizing the emission shell
\citep{nak07}. 
The decay times of short GRBs are shorter than those of the long ones. 
Assuming an internal shock origin for both, we consider the possible 
implications of our observational results. If the curvatures of the emitting 
shells ($R_e$) are similar for both GRB types, then shorter decay times would 
imply that the $\gamma-$ray emitting shells of short bursts have significantly 
larger Lorentz factors. On the other hand if the Lorentz factors are similar 
then the radii of the emitting shells are smaller in short GRBs, implying a 
more compact central engine. Ackermann et al. (2010) compare the estimates of 
$\Gamma_{\rm min}$ (the bulk Lorentz factors) for two long and one short GRBs. 
These are 900, 1000 and 1218, respectively, possibly indicating 
(albeit with small number statistics) that the shell radii of short 
GRBs are significantly smaller. 

According to \citet{der09} the GRB central engine releases energy at a fixed 
rate over a time scale $\Delta _0/c$, where $\Delta _0$ is a characteristic 
size scale of the engine. Assuming that the shortest time scale in GRB prompt 
emission is the shortest pulse width, we can estimate the length scale of the 
GRB central engine. Using the mean shortest FWHM of short and long GRBs, 
0.016\,s and 0.087\,s, respectively, we find their corresponding length scales 
to be $4.8\times10^8$\,cm and $2.6\times10^{9}$\,cm, respectively in the 
observer frame. To convert these length scales to the source frame, we use the 
mean redshifts, $z=2.245, 0.862$, of 151 long and 12 short {\it Swift} GRBs
\footnote{\url{http://swift.gsfc.nasa.gov/docs/swift/archive/grb\_table/}}. 
These are $8\times 10^8$ cm and $2.6\times 10^8$ cm, respectively. According 
to the GRB standard model \citep{mes06} the above length scales agree with the 
saturation radius of the fireball ($\sim 10^{9-10}$\,cm), $i.e.,$ the radius 
signifying the end of the acceleration phase and the beginning of the coasting 
phase of the Lorentz factor $\gamma_e$. Our results imply that the central 
engines of short GRBs seem to have a relatively smaller saturation radii. 

We now estimate the rest frame radii of the shells, which give rise to the pulses. 
According to \cite{der04} the radius of the emission shell $R_e$ is given by:
\begin{displaymath}
R_e \approx \frac{2\gamma_e ^2 ct_{var}}{1+z}
\end{displaymath}  
where $t_{var}$ is the GRB light curve variability time scale. We substitute 
$t_{var}$ with the mean FWHM values of the GRB pulses (which are 0.9 and 0.06 s
for long and short bursts respectively) and assuming a typical 
value for $\Gamma_e \approx 1000$, we find that the mean shell radii are 
$1.7\times 10^{16}$\,cm and $1.9\times 10^{15}$\,cm for long and short bursts, 
respectively. \cite{zha11} find shell radii for the long GRB\,080916C that are 
slightly larger but comparable to the above mean value. Even larger prompt 
emission radii were inferred for other GRBs by different estimates 
\citep{kum07,rac08}. Our mean shell radii agree with the radial distances when 
the internal shock phase ($\sim 10^{14-15}$\,cm) or the prompt emission starts
\citep{mes06}, possibly 
indicating that the beginning of the internal shock phase occurs earlier for 
short bursts. 

If the individual pulses in the GRB light curves are indeed formed by the 
collision of shells with unequal Lorentz factors \citep{rmm94,enp02} then 
shorter intervals between pulses (Figure \ref{GRB_int}) imply that the 
relativistic shells are more frequent. However, the longer intervals between successive pulses and 
durations of long GRBs indicate that the central engine shell ejection 
persists for longer times. In other words, the duration as well as the 
structure of the light curve are indeed related to the central engine activity. 

Temporal analysis of long and short GRB light curves carried out here supports
the general observation that the short bursts are temporally similar to
long ones but compressed in time, which could be related to the nature of the
central engine of the respective bursts. 

\section{Acknowledgments}

The GBM project is supported by the German Bundesministerium f¬ur 
Wirtschaft und Technologie (BMWi) via the Deutsches Zentrum f¬ur Luft-und 
Raumfahrt (DLR) under the contract numbers 50 QV 0301 and 50 OG 0502.

AJvdH was supported by NASA grant NNH07ZDA001-GLAST.

SMB acknowledges support of the Union Marie Curie European 
Reintegration Grant within the 7th Program under contract number 
PERG04-GA-2008-239176.

SF acknowledges the support of the Irish Research Council for Science, 
Engineering and Technology, cofunded by Marie Curie Actions under FP7.

We also acknowledge the constructive comments and suggestions from the anaonymous
referee which improved the quality of presentation.


\begin{deluxetable}{rrr}
\tablecolumns{3}
\tablewidth{0pc}
\tablecaption{List of GRBs, chosen for the present analysis. Also listed are 
their durations (column 2) and the number of fitted pulses (column3) 
\label{GRB_table}}
\tablehead{
\colhead{Burst $\#$} & \colhead{Duration $T_{90}$ (s)}   & \colhead{$\#$ of Fitted Pulses}}
\startdata
\cutinhead{Long GRBs}
bn080723557  &  $58.37\pm 1.98$  &    29  \\
bn080723985  &  $42.80\pm 0.66$  &    18  \\
bn080807993  &  $19.07\pm 0.18$  &    16  \\
bn080817161  &  $60.29\pm 0.47$  &    14  \\
bn080825593  &  $20.99\pm 0.23$  &    20  \\
bn080906212  &  $2.875\pm 0.77$  &     5  \\
bn080916009  &  $62.98\pm 0.81$  &    32  \\
bn080925775  &  $31.74\pm 3.17$  &    15  \\
bn081009690  &  $176.2\pm 2.13$  &     5  \\
bn081101532  &  $8.260\pm 0.90$  &     7  \\
bn081110601  &  $17.34\pm 0.68$  &     2  \\
bn081121858  &  $41.98\pm 8.51$  &     9  \\
bn081122520  &  $23.30\pm 2.11$  &     6  \\
bn081125496  &  $9.280\pm 0.61$  &     4  \\
bn081129161  &  $62.66\pm 7.32$  &     2  \\
bn081207680  &  $97.28\pm 2.35$  &     3  \\
bn081215784  &  $5.570\pm 0.14$  &    10  \\
bn081224887  &  $16.45\pm 1.16$  &     4  \\
bn081231140  &  $28.74\pm 2.61$  &     3  \\
bn090102122  &  $26.62\pm 0.81$  &    25  \\
bn090131090  &  $35.07\pm 1.06$  &     8  \\
bn090217206  &  $33.28\pm 0.72$  &    23  \\
bn090323002  &  $135.2\pm 1.45$  &    17  \\
bn090328401  &  $61.70\pm$ 1.81  &     7  \\
bn090424592  &  $14.14\pm$ 0.26  &    15  \\
bn090425377  &  $75.39\pm$ 2.45  &     3  \\
bn090528516  &  $79.04\pm$ 1.09  &    16  \\
bn090529564  &  $9.850\pm$ 0.18  &     7  \\
bn090618353  &  $112.4\pm$ 1.09  &    20  \\
bn090620400  &  $13.57\pm$ 0.72  &     5  \\
bn090623107  &  $47.11\pm$ 2.57  &    18  \\
bn090626189  &  $48.90\pm$ 2.83  &    24  \\
\cutinhead{Short GRBs}
    bn080905499 &   $ 0.960\pm $ 0.35 &    7  \\
    bn081209981 &   $ 0.192\pm $ 0.14 &    2  \\
    bn081216531 &   $ 0.768\pm $ 0.43 &    7  \\
    bn090227772 &   $ 1.280\pm $ 1.03 &    5  \\
    bn090228204 &   $ 0.448\pm $ 0.14 &    9  \\
    bn090305052 &   $ 1.856\pm $ 0.58 &   10  \\
    bn090308734 &   $ 1.664\pm $ 0.29 &    7  \\
    bn090429753 &   $ 0.640\pm $ 0.47 &    3  \\
    bn090510016 &   $ 0.960\pm $ 0.14 &   12  \\
    bn090617208 &   $ 0.192\pm $ 0.14 &    3  \\
\enddata
\end{deluxetable}

\begin{table}
\begin{center}
\caption{The lower edges of the 8 energy channels used for pulse fitting of GRB
 light curves using the 2 types of GBM detectors. The upper energy edge of 
channel 7 is assumed to be twice the lower energy edge of that channel. 
\label{Tab1}}
\begin{tabular}{crrrrrrrrrrr}
\tableline
Channel $\#$ & 0 & 1  & 2 & 3 & 4 & 5  & 6  & 7 \\
\tableline
NaI (keV) & 8.0 & 20 & 40 & 70 & 142 & 270 & 524 & $>$985 \\
\tableline
BGO (MeV) & 0.11 & 0.28 & 0.55 & 1.4 & 3.3 & 7.2 & 19.2 & $>$45.5 \\
\tableline
\end{tabular}
\end{center}
\end{table}
 
\begin{table}
\begin{center}
\caption{The variation of the number of fitted pulses in various NaI energy
ranges for a sample of long bursts.\label{Tab3}}
\begin{tabular}{crrrrrrrrrrr}
\tableline
NaI Energy Range (keV) & 8-20 & 20-40 & 40-70 & 70-142 & 142-270 & 270-524  \\
Mean Energy (keV) & 12.5 & 28.3  & 53.5  & 100.0  & 195.6   &  376.1   \\
\tableline
bn081207680 &2    &      2      &    2     &     2     &     2   &       2\\
bn081215784 &8    &      9      &    9     &     9     &     8   &       8\\
bn081231140 &3    &      3      &    3     &     3      &    3   &       2\\
bn090217206 &19   &      20    &     23    &     26    &     19  &       19\\
bn090323002 &18   &      16    &     20    &     26    &     20  &       18\\
bn090328401 &6    &      6     &     4     &     5     &     6   &       5\\
bn090424592 &13   &      15    &     16    &     17    &     14  &       11\\
bn090529564 &7    &      8     &     6     &     9     &     8   &       6\\
bn090618353 &23   &      22    &     24    &     26    &     30  &       24\\
bn090626189 & 30  & 33          & 32       & 32         & 26     & 20    \\
\tableline
\end{tabular}
\end{center}
\end{table}

\end{document}